\documentstyle[preprint,aps,epsf]{revtex}
\begin{document}

\title{Non perturbative chiral approach to s-wave $\bar{K} N$
interactions}

\author{E. Oset}

\address{
Departamento de F\'{\i}sica Te\'orica and IFIC,
Centro Mixto Universidad de Valencia-CSIC, \\
46100 Burjassot (Valencia), Spain}

\author{A. Ramos}
\address{Departament d'Estructura i Constituents de la Mat\`eria,
Universitat de Barcelona, \\
Diagonal 647, 08028 Barcelona, Spain
}

\date{\today}

\maketitle
\begin{abstract}
The s-wave meson-nucleon interaction in the $S = -1$ sector is studied by
means of coupled-channel Lippmann Schwinger equations, using the lowest order
chiral Lagrangian and a cut off to regularize the loop integrals. The method
reproduces succesfully the $\Lambda (1405)$ resonance and the
$K^- p \rightarrow K^- p, \bar{K}^0 n, \pi^0 \Lambda, \pi^0 \Sigma ,
\pi^+ \Sigma^-, \pi^- \Sigma^+$ cross sections at low energies. The inclusion
of the $\eta \Lambda, \eta \Sigma^0$ channels in the coupled system is found
very important and allows a solution in terms of only the lowest
order Lagrangian.

\end{abstract}

\pacs{12.39.Fe, 13.75.Jz}

\section{Introduction}

The effective chiral Lagrangian formalism which has proved successful in
explaining the properties of meson-meson interaction at low energies
\cite{Ga85,Mei93,Pi95} has also proved to be an idoneous tool to study
low energy properties of the meson-baryon interaction \cite{Eck95,Be95}.
The s-wave $\pi N$ and $K^+ N$ interaction is relatively weak and the leading
term in the chiral expansion $O (q)$ is the dominant one \cite{Lee94,Bro93}.
By contrary, in the $S = -1$ sector, the $\bar{K} N$ system couples strongly
to many other channels and generates a resonance below threshold in s-wave,
the $\Lambda (1405)$. In such case the standard chiral perturbative scheme,
an expansion in powers of the typical momenta involved in the process, fails
to be an appropiate approach, since the singularities of the $T$ matrix
associated to the resonance cannot be generated perturbatively.

A non perturbative scheme to the $S = -1$ meson-baryon sector, yet using the
input of the Chiral Lagrangians, was employed in \cite{Kai95}. A set of
coupled-channel Lippmann Schwinger (LS) equations
was solved using the lowest and next
to lowest order chiral Lagrangians. The $\Lambda (1405)$ resonance was
generated and the cross sections of the
 $K^- p \rightarrow K^- p, \bar{K}^0 n, \pi^0 \Lambda, \pi^+ \Sigma^-,
\pi^0 \Sigma^0, \pi^- \Sigma^+$ reactions at low energies, plus the threshold
branching ratios, were well reproduced. In summary, five
parameters were needed to fit the experimental information, corresponding to,
so far, unknown parameters of the second order chiral Lagrangian plus some
range parameters used to construct a potential from the chiral Lagrangians.
The method was also used to study coupled channels in the $\pi N$ sector plus
eta meson and kaon photoproduction in \cite{Kai97}.
The role of the resonance is so important in
$\bar{K}N$ scattering at low energies
that any finite-order chiral expansion will fail to
reproduce the data, unless the $\Lambda(1405)$ is introduced as an
elementary matter field\cite{CHL96}.
Another approach based on the coupled-channel LS equations \cite{Sie95}
started from transition potentials
whose relative strength between various channels was guided by SU(3)
symmetry but was allowed to be broken by up to $\pm 50\%$ in order to fit the
data.

The success of Ref. \cite{Kai95} has stimulated work in the
meson-meson sector. In \cite{OO97} similar ideas were followed and, by means
of coupled-channel LS equations using the lowest order chiral Lagrangian,
plus a suitable cut off in the loops in order to simulate the effect of the
second order Lagrangian, an excellent reproduction of the $\sigma, f_0 (980),
a_0 (980)$ resonances in the scalar sector, plus phase shifts and
inelasticities in the different physical channels was obtained. The work
required just one free parameter, the cut off, $q_{max}$, in the momentum
of the loop. However, the extension of these ideas to the $L = 1$ sector
proved that the cut off alone was insufficient to account for the information
contained in the second order chiral Lagrangians and the generation of the
$\rho$ and $K^*$ resonances required further input.

In \cite{OOP97} the method of \cite{OO97} was generalized using
ideas of the
inverse amplitude method \cite{DHT90,DP92} leading to a unitary
coupled-channel
non perturbative scheme that includes the works of \cite{OO97} and
\cite{DHT90,DP92} as particular cases. It uses the input of the first and
second order Lagrangians and a cut off regularization and reproduces all the
meson-meson experimental information up to $\sqrt{s} = 1.2$ GeV,
including
the resonances $\sigma, f_0 (980), a_0 (980), \rho$ and $K^*$. The work
requires the use of 7 parameters, coefficients of the second order chiral
Lagrangians in the meson-meson interaction \cite{Ga85}.

In the present work we want to extend the ideas of \cite{OO97} to the
$\bar{K} N$ sector and investigate the possibility to describe all the
low
energy experimental cross sections plus the $\Lambda (1405)$ resonance in
terms of the  lowest order chiral Lagrangian (with no free parameters) and
one cut off. As we shall see, we succeed in the enterprise, thus stressing
the role of chiral symmetry in the meson-baryon interaction and at the same
time the usefulness of the unitary coupled-channel method of \cite{OO97} to
deal with  this kind of reactions.

The work presented here shares many points with \cite{Kai95} but has one
different main result. The authors of \cite{Kai95} were able to reproduce
fairly well the experimental
cross sections with just the lowest order Lagrangian, but found substantial
differences with the threshold branching ratios.
We can reproduce all the results with the lowest order Lagrangian and one
cut off.  The main reason
for the differences is the inclusion of two extra channels in our approach.
In \cite{Kai95} the $K^- p, \bar{K}^0 n, \pi^0 \Lambda, \pi^+ \Sigma^-,
\pi^0 \Sigma^0$ and $\pi^- \Sigma^+$ channels were considered. The
$\eta
\Lambda$
and $\eta \Sigma^0$ channels open up at higher $K^-$ energies than studied
in \cite{Kai95} and thus they were omitted in that work. We have included these
channels in our approach using the analytical extrapolation of these
amplitudes below threshold and find substantial effects in the cross
sections, changing the key threshold ratios in more than a factor two.

\section{Meson-nucleon amplitudes to lowest order}

Following \cite{Pi95,Eck95,Be95} we write the lowest order chiral Lagrangian,
coupling the octet of pseudoscalar mesons to the octet of $1/2^+$ baryons, as

$$
L_1^{(B)} = < \bar{B} i \gamma^{\mu} \nabla_{\mu} B> - M_B <\bar{B} B> +
$$

\begin{equation}
\frac{1}{2} D <\bar{B} \gamma^{\mu} \gamma_5 \left\{ u_{\mu}, B \right\} >
+ \frac{1}{2} F <\bar{B} \gamma^{\mu} \gamma_5 [u_{\mu}, B]>
\end{equation}

where the symbol $< \, >$ denotes trace of SU(3) matrices and

\begin{equation}
\begin{array}{l}
\nabla_{\mu} B = \partial_{\mu} B + [\Gamma_{\mu}, B] \\
\Gamma_{\mu} = \frac{1}{2} (u^+ \partial_{\mu} u + u \partial_{\mu} u^+) \\
U = u^2 = {\rm exp} (i \sqrt{2} \Phi / f) \\
u_{\mu} = i u ^+ \partial_{\mu} U u^+
\end{array}
\end{equation}

The SU(3) matrices for the mesons and the baryons are the following

\begin{equation}
\Phi =
\left(
\begin{array}{ccc}
\frac{1}{\sqrt{2}} \pi^0 + \frac{1}{\sqrt{6}} \eta & \pi^+ & K^+ \\
\pi^- & - \frac{1}{\sqrt{2}} \pi^0 + \frac{1}{\sqrt{6}} \eta & K^0 \\
K^- & \bar{K}^0 & - \frac{2}{\sqrt{6}} \eta
\end{array}
\right)
\end{equation}

\begin{equation}
B =
\left(
\begin{array}{ccc}
\frac{1}{\sqrt{2}} \Sigma^0 + \frac{1}{\sqrt{6}} \Lambda &
\Sigma^+ & p \\
\Sigma^- & - \frac{1}{\sqrt{2}} \Sigma^0 + \frac{1}{\sqrt{6}} \Lambda & n \\
\Xi^- & \Xi^0 & - \frac{2}{\sqrt{6}} \Lambda
\end{array}
\right)
\end{equation}

At lowest order in momentum, that we will keep in our study, the interaction
Lagrangian comes from the $\Gamma_{\mu}$ term in the covariant derivative
and we find

\begin{equation}
L_1^{(B)} = < \bar{B} i \gamma^{\mu} \frac{1}{4 f^2}
[(\Phi \partial_{\mu} \Phi - \partial_{\mu} \Phi \Phi) B
- B (\Phi \partial_{\mu} \Phi - \partial_{\mu} \Phi \Phi)] >
\end{equation}

\noindent
which leads to a common structure of the type
$\bar{u} \gamma^u (k_{\mu} + k'_{\mu}) u$ for the different channels, where
$u, \bar{u}$ are the Dirac spinors and $k, k'$ the momenta of the incoming
and outgoing mesons.

We take the $K^- p$ state and all those that couple to it within the chiral
scheme. These states are $\bar{K}^0 n, \pi^0 \Lambda, \pi^0 \Sigma^0,
\pi^+ \Sigma^-, \pi^- \Sigma^+, \eta \Lambda, \eta \Sigma^0$. Hence we have
a problem with eight coupled channels. We should notice that, in addition to
the six channels considered in \cite{Kai95} we have the two $\eta$ channels,
$\eta \Lambda$ and $\eta \Sigma^0$. Although these channels are above
threshold for $K^- p$ scattering at low energies, they couple strongly to the
$K^- p$ system and there are important interferences between the real parts
of the amplitudes, which make their inclusion in the coupled-channel approach
very important as we shall see.

The lowest order amplitudes for these channels are easily evaluated from eq.
(5) and are given by

\begin{equation}
V_{i j} = - C_{i j} \frac{1}{4 f^2} \bar{u} (p') \gamma^{\mu} u (p)
(k_{\mu} + k'_{\mu})
\end{equation}

\noindent
were $p, p' (k, k')$ are the initial, final momenta of the baryons (mesons).
Also, for low energies one can safely neglect the spatial components in eq.
(6) and only the $\gamma^0$ component becomes relevant, hence simplifying
eq. (6) which becomes

\begin{equation}
V_{i j} = - C_{i j} \frac{1}{4 f^2} (k^0 + k'^0)
\end{equation}

The matrix $C_{i j}$, which is symmetric, is given in Table I.

\section{Isospin formalism}

We shall construct the amplitudes using the isospin formalism for
which we must use average masses for the $K$ ($K^-, \bar{K}^0$), $N$
$(p,n)$, $\pi$ ($\pi^+,\pi^0,\pi^-$) and $\Sigma$
($\Sigma^+,\Sigma^0,\Sigma^-$) states.
The isospin amplitudes are

\begin{eqnarray}
\mid \bar{K} N, T = 0 \rangle & =&  \frac{1}{\sqrt{2}} (\bar{K}^0 n +
K^- p)
\nonumber \\
\mid \bar{K} N, T = 1\rangle & =&  \frac{1}{\sqrt{2}} (\bar{K}^0 n -
K^- p)
\\
\mid \pi \Sigma, T = 0\rangle & =&  - \frac{1}{\sqrt{3}}
( \pi^+ \Sigma^- + \pi^0 \Sigma^0 + \pi^- \Sigma^+) \nonumber \\
\mid \pi \Sigma, T = 1 \rangle & =&  \frac{1}{\sqrt{2}}
( \pi^- \Sigma^+ - \pi^+ \Sigma^-)  \nonumber \ ,
\end{eqnarray}
where we use the phase convention $\mid \pi^+ \rangle = - \mid 1, 1
\rangle$, $\mid \Sigma^+ \rangle = - \mid 1, 1 \rangle$,
$\mid~K^-~\rangle~=~-~\mid~1/2,~-1/2~\rangle $ for the isospin states,
consistent with the structure of the $\Phi$ and $B$ matrices.

In $T = 0$ we have three channels, $\bar{K} N, \pi \Sigma$ and $\eta \Lambda$
while in $T = 1$ we have four channels,
$\bar{K} N, \pi \Sigma, \pi \Lambda, \eta \Sigma$. Using eqs. (8) and Table
I we can construct the transition matrix elements in isospin formalism which
read

\begin{eqnarray}
V_{i j} (T = 0) &=& - D_{i j} \, \frac{1}{4 f^2} \, (k^0 + k'^0)
\nonumber \\
V_{i j} (T = 1) &=& - F_{i j} \, \frac{1}{4 f^2} \, (k^0 + k'^0)
\end{eqnarray}
and the symmetrical $D_{i j}, F_{i j}$ coefficients are given in Tables II
and III.

An alternative treatment can be done using directly the
physical channels and physical masses of the particles. We
shall make use of it too in order to investigate the isospin violation
effects.

\section{Amplitudes in other strangeness and isospin channels.}

For completeness we give here the $S = - 1, T = 2$ and $S = 1$ channels,
plus the $S = - 1$ in $K^- n$ and related channels.

\vspace{0.5cm}

a) $S = - 1, T = 2$ channel:

\vspace{0.5cm}

Only the $\pi \Sigma$ state couples to this channel. We take
$\mid \pi^+ \Sigma^+ \rangle \equiv \mid 2, 2\rangle$
and the potential in this case is given by

\begin{equation}
V = \frac{1}{2 f^2} \, (k^0 + k'^0)
\end{equation}

\vspace{0.5cm}

b) $S = 1$ channel.

\vspace{0.5cm}

We take $K^+ n$ and the coupled state $K^0 p$, which are admixtures of
$T = 0, T = 1$.
The potential in this case is given by

\begin{equation}
V_{i j} = - \frac{1}{4 f^2} \; \; L_{i j} (k^0 + k'^0)
\end{equation}

\noindent
with the $L_{i j}$ coefficients given in Table IV.

The $K^+ p$ state stands alone for the $T = 1, T_3 = 1$ channel. The
potential is given by

\begin{equation}
V = \frac{1}{2 f^2} \; (k^0 + k'^0)
\end{equation}

The isospin amplitudes are written immediately and we have

\begin{eqnarray}
V (S = 1, T = 0) &=& 0 \nonumber \\
V (S = 1, T = 1) &=& \frac{1}{2 f^2} (k^0 + k'^0)
\end{eqnarray}

As we can see, at lowest order the $S = 1, T = 0$ amplitude vanishes.
When working with the physical
masses of the $K^+$,$K^0$, $p$ and $n$, the coupling of the channels
breaks slightly this symmetry
but still leads to a very small amplitude as
we shall see.

\vspace{0.5cm}

c)$S = -1, K^- n$ and related channels.

\vspace{0.5cm}

For the purpose of $K^-$ nucleus interaction we shall also need the $K^- n$
amplitude which we evaluate here. The coupled channels in this case, which
is only $T = 1$, are $K^- n, \pi^0 \Sigma^-, \pi^- \Sigma^0, \pi^- \Lambda,
\eta \Sigma^-$. Since the matrix elements of the potential satisfy isospin
symmetry, these matrix elements are easily induced from section 3 and
Tables II and III. We have

\begin{equation}
V_{i j}= - \tilde{C}_{i j} \, \frac{1}{4 f^2} \, (k^0 + k'^0)
\end{equation}

\noindent
where the $\tilde{C}_{i j}$ coefficients are given in Table V.

\section{Coupled channels Lippmann Schwinger equations}

Following \cite{OO97} we write the set of Lippmann Schwinger equations in
the $\bar{K} N$ centre of mass frame

\begin{equation}
t_{i j} = V_{i j} + V_{i l} \; G_l \; T_{l j}
\end{equation}

\noindent
where the indices $i,j$ run over all possible channels and

\begin{equation}
V_{i l} \; G_l \; T_{l j} = i \int \frac{d^4 q}{(2 \pi)^4} \,
\frac{M_l}{E_l (\vec{q})}
\, \frac{V_{i l} (k, q) \, T_{l j} (q, k')}{k^0 + p^0 - q^0 - E_l (\vec{q})
+ i \epsilon} \, \frac{1}{q^2 - m^2_l + i \epsilon}
\end{equation}

Eq. (15) sums up automatically the series of diagrams of fig. 1. In
eq. (16)
we have kept the positive energy part of the baryon propagator, although
with proper relativistic factors in order to ensure exact phase space in the
imaginary part of the expressions. In eq. (16) $M_l, E_l$ correspond
to the
mass and energy of the intermediate baryon and $m_l$ to the mass of the
intermediate meson.

The integral of eq. (16) is regularized through the use of a
momentum cut off,
$q_{max}$.
The value of $q_{max}$ is a free parameter
of the theory by means of which one accounts for higher order contributions
in an effective way.

Some other comments must be made with respect to the off shell extrapolation
of $V_{i l} (k, q)$ which run in parallel to the findings of
\cite{OO97}. In
that work the potential was split into an on shell part plus a rest.
The contribution from
this latter part was found to go into renormalization of couplings and
masses and could hence be omitted in the calculation. This simplified the
coupled integral equations which became then ordinary algebraic
equations. The same happens here, as we see below.

Let us take the one loop diagram of fig. 1 and equal masses in the external
and intermediate states for simplicity. We have

\begin{equation}
\begin{array}{l}
V^2_{off} = C (k^0 + q^0)^2 = C (2 k^0 + q^0 - k^0)^2 \\
\hspace{1cm}
= C^2 (2 k^0)^2 + 2 C (2 k^0) (q^0 - k^0) + C^2 (q^0 - k^0)^2
\end{array}
\end{equation}

\noindent
with $C$ a constant. The first term in the last expression is the on shell
contribution $V^2_{on} (V_{on} \equiv C 2 k^0)$. Neglecting $p^0 - E (q)$
in eq. (16), typical approximations in the heavy baryon formalism
\cite{JM91}, the one loop integral for the second term of eq. (17)
becomes
$(\omega (q)^2 = \vec{q} \, ^2 + m^2)$

\begin{eqnarray}
2 i V_{on} \int \frac{d^3 q}{(2 \pi)^3} \, \int \, \frac{d q^0}{2 \pi} \,
\frac{M}{E (q)} \, \frac{q^0 - k^0}{k^0 - q^0} \,
\frac{1}{q^{02} - \omega (q)^2 + i \epsilon} = \nonumber \\
- 2 V_{on} \, \int \, \frac{d^3 q}{(2 \pi)^3} \, \frac{M}{E (q)} \,
\frac{1}{2 \omega (q)} \sim V_{on} \, q^2_{max}
\end{eqnarray}

As we can see this term is proportional to $V_{on}$ and hence can be reabsorbed
by a suitable renormalization of the coupling $f$. Therefore, the use
of the physical coupling will incorporate this term. In the case of
coupled channels the arguments are similar. The contribution of eq. (18)
has the same structure as the lowest order terms and can be reabsorbed in
the lowest order Lagrangian by a suitable renormalization, leading to the
effective chiral Lagrangian with the physical couplings.

Similarly, the term proportional to $(q^0 - k^0)^2$ will cancell the
$(k^0 - q^0)$ term in the denominator and the integral of this term,
proportional to
$(k^0 - q^0)$, gives rise to another term proportional to $k^0$ (and hence
$V_{on}$) while the term proportional to $q^0$ vanishes for parity
reasons.

We can extend these arguments to higher order loops and the conclusion is that
we can factorize $V_{on}$ and $T_{on}$ outside the integral of
eq. (16). Hence in matrix form we will have

\begin{equation}
T = V + V \, G \, T
\end{equation}

\noindent
or equivalently

\begin{equation}
T = [1 - V \, G]^{-1}\, V
\end{equation}

\noindent
with $G$ a diagonal matrix given by

\begin{eqnarray}
G_{l} &=& i \, \int \frac{d^4 q}{(2 \pi)^4} \, \frac{M_l}{E_l
(\vec{q})} \,
\frac{1}{k^0 + p^0 - q^0 - E_l (\vec{q}) + i \epsilon} \,
\frac{1}{q^2 - m^2_l + i \epsilon} \nonumber \\
&=& \int \, \frac{d^3 q}{(2 \pi)^3} \, \frac{1}{2 \omega_l
(q)}
\,
\frac{M_l}{E_l (\vec{q})} \,
\frac{1}{p^0 + k^0 - \omega_l (\vec{q}) - E_l (\vec{q}) + i \epsilon}
\end{eqnarray}

\noindent
which depends on $p^0 + k^0 = \sqrt{s}$ and $q_{max}$.

The method of \cite{OOP97} provides an alternative reinterpretation of the
on shell factorization which is clarifying. The method uses the optical
theorem to start with, which is stated here as

\begin{equation}
Im\, T = T \, Im \, G \, T^*
\end{equation}

\noindent
from where one deduces
\begin{equation}
Im\, G = - Im \, T^{-1}
\end{equation}

Hence
\begin{equation}
\begin{array}{l}
T = [Re \, T^{-1} - i \, Im \, G]^{-1} = \\
V [V \, Re \, T^{-1} \, V - i \, V \, Im \, G \, V]^{-1} \, V
\end{array}
\end{equation}

\noindent
where in the last step we have multiplied twice by $V V^{-1}$ for convenience,
with $V \equiv V_{on}$. Expanding formally $V\, Re\,T^{-1}\, V$ in
powers of a
suitable parameter, proportional to $k^0$ for instance, one obtains up to
$2^{nd}$ order

\begin{equation}
T = V\, [V - Re\, T_2 - i\, V\, Im\, G\, V]^{-1}\, V
\end{equation}

\noindent
with $T_2$ the second term in the expansion of $T (T= T_1 + T_2,\ T_1
\equiv V)$.
The freedom of the cut off can be used to make $Re\, T_2 \simeq V\,
Re\, G\, V$, in
which case eq. (25) reduces to the LS equations implicit in eq. (19).
The success of the
LS method in \cite{OO97} suggests that the expansion of $V\, Re\,
T^{-1}\, V$,
and the approximation to $Re\, T_2$ given above, are sensible
approximations
at least in the scalar sector for the meson-meson interaction. One
hopes
that this is also the case for the meson-baryon interaction in $L = 0$ that
we study here.

The coupled-channel equations represented by eq. (19)
are solved in the isospin basis for the $T = 0,\ T = 1$ cases,
from where the amplitudes in the physical channels are then constructed.
Alternatively we can work directly with the physical states using the matrix
of Table I and the physical masses of each particle. The second method is
more accurate and respects exactly the thresholds for the reaction and the
phase space. We use both methods and this allows us to see the amount of
isospin violation in the different channels.

The channels $\eta \Lambda, \eta \Sigma$ are above threshold for low energies
of the $K^-$. The potential $V_{i j} (s)$ for these channels is taken through
an analytical continuation using the formula

\begin{equation}
k^0 = \frac{s + m_{\eta}^2 - M_B^2}{2 \sqrt{s}}
\end{equation}

\section{The  ${\mbox{\bf $\Lambda$}}$ (1405) resonance and the
${\mbox{\bf $\pi \Sigma$}}$ mass spectrum.}

The $\Lambda (1405)$ resonance appears below the $K^- p$ threshold. It is
observed in the mass spectrum of $\pi \Sigma$. One of the reactions used
to see it is $\pi^- p \rightarrow K^0 \Sigma^+ \pi^-$ \cite{Th73}.

According to \cite{Fl76}, the mass distribution of the $\Sigma^+ \pi^-$
state, for s-wave resonance, is given by

\begin{equation}
\frac{d \sigma}{dm_{\alpha}} = C |t_{\pi \Sigma \rightarrow \pi \Sigma}|^2
p_{CM}
\end{equation}

\noindent
where $C$ is a constant, $t_{\pi \Sigma \rightarrow \pi \Sigma}$ is
the $T = 0 \;
\pi \Sigma$ amplitude and $p_{CM}$ is the $\pi$ momentum in the frame where
$\pi \Sigma$ is at rest.

\section{Results}

With the normalization which we use, the cross section is given by

\begin{equation}
\sigma_{i j} = \frac{1}{4 \pi} \, \frac{M M'}{s} \,
\frac{k'}{k} \, |T_{i j}|^2
\end{equation}

The relationship to the scattering lengths in elastic channels reported in
\cite{Kai95} is

\begin{equation}
a_i = - \frac{1}{4 \pi} \, \frac{M}{\sqrt{s}} \, T_{i i}
\end{equation}

\noindent
calculated at threshold.

We look at the cross sections for
$K^- p \rightarrow K^- p, \bar{K}^0 n, \pi^0 \Lambda, \pi^0 \Sigma^0,
\pi^+ \Sigma^-, \pi^- \Sigma^+$ at low energies plus the $\pi \Sigma$
mass distribution and the threshold branching ratios.
Our free parameter is $q_{max}$, but we allow also
some small variation of $f$ from the pionic value of $f_{\pi} = 93$ MeV.
For kaons in the meson-meson interaction $f_K = 1.22 f_{\pi}$ and we
should expect a similar renormalization here. However, for simplicity, we use
a single value of $f$ for pions and kaons which is fit to the data and turns
out to be between $f_\pi$ and $f_K$.

The threshold branching ratios which we
use in the fitting procedure, as in \cite{Kai95}, are \cite{No78,To71}:

\begin{eqnarray}
\gamma &=& \frac{\Gamma (K^- p \rightarrow \pi^+ \Sigma^-)}
{\Gamma (K^- p \rightarrow \pi^- \Sigma^+)} = 2.36 \pm 0.04
\nonumber \\
R_c &=& \frac{\Gamma (K^- p \rightarrow \hbox{charged particles)}}
{\Gamma (K^- p \rightarrow \hbox{all)}} = 0.664 \pm 0.011 \\
R_n &=& \frac{\Gamma (K^- p \rightarrow \pi^0 \Lambda)}
{\Gamma (K^- p \rightarrow \hbox{all neutral states})} = 0.189 \pm 0.015
\nonumber
\end{eqnarray}

Note that the ratio $\gamma$ is zero in lowest order of the chiral Lagrangians
(see Table I). The coupled-channel LS equations lead to a finite cross
section for $K^- p \rightarrow \pi^+ \Sigma^-$ which is larger than the
$K^- p \rightarrow \pi^- \Sigma^+$ as we shall see.

Our fitting procedure proceeds as follows: first we fix a value of $f$
around $f_{\pi} = 93$ MeV and vary $q_{max}$ in order to get the
best reproduction of the threshold parameters, $\gamma, R_c, R_n$.
There is a correlation between the values of $q_{max}$ and $f$ leading to
the best fit to these threshold parameters. A $2 \%$ increase in $f$ can be
compensated with a $3 \%$ increase in $q_{max}$. The shape and position of
the $\Lambda (1405)$ resonance depend on the value of $f$ (and its associated
$q_{max}$ from the previous fit) and we choose the value of $f$ which leads
to the best agreement with the $\Lambda (1405)$ properties seen in the
$\pi \Sigma$ mass spectrum.
This procedure determines
$f, q_{max}$ and no further input is used in the fit. The cross sections
are then calculated with the best choice of parameters and have not
not been used in a best
fit to the data. As we shall see, it is a remarkable feature of this chiral
coupled-channel approach that the threshold ratios plus the position and
shape of the $\Lambda (1405)$ determine the behaviour of the $K^- p$ cross
sections at low energies in all channels.

Our optimal choice was found for $f = 1.15 f_{\pi}, q_{max} = 630$
MeV. The
following results are evaluated inverting the $8 \times 8$ matrix
$(1 - V\, G)$ with $V$ given in Table I. We will also show the results
obtained
using the isospin basis and inverting $(1 - V\, G)$ with V given by
Tables II and
III. At the same time we show the results obtained omitting the
$\eta \Lambda$ and $\eta \Sigma^0$ channels as done in \cite{Kai95}.

In fig. 2 we show the $\pi \Sigma$ spectrum corresponding to the $\Lambda
(1405)$ resonance. As we can see, the peak position and width are well
reproduced. The results obtained using the
isospin basis and those omitting the $\eta$ channels are also shown
in the figure. The results with
the isospin basis are similar to those obtained with the basis of physical
states, however, omitting the $\eta$ channels leads to a quite
different mass distribution, which is
incompatible with the data. Obviously one can choose other
values of $f$ and $q_{max}$ to reproduce the mass distribution without
the $\eta$ channels but,
as shown in \cite{Kai95} and corroborated here, one can not obtain a
global fit to the data. In any case, one of the points in this paper
is to
show the relevance of the inclusion of the $\eta$ channels in the
coupled-channel
equations, and the results for the $\Lambda (1405)$ resonance are a clear
example of it, although more spectacular effects on other
observables will be shown in the following.

In Table VI we display the results for the threshold ratios evaluated
in the
three cases: isospin basis, full basis and omitting $\eta$ channels.
We can see that the three ratios are reproduced within
$5 \%$ in the calculation with the full basis.
Note that using the isospin basis or omitting the $\eta$
channels produces appreciable changes in these ratios. Particularly
remarkable
is the change in the ratio $\gamma$, which is reduced by a factor 2.2
when the $\eta$ channels are omitted.
It is worth mentioning that the small values for $\gamma$ obtained in
\cite{Kai95}, which are compatible with our value when
the $\eta$ channels are omitted, motivated the authors of that work to
introduce
higher order terms in the chiral expansion and perform a global fit with
five parameters.

In figs. 3--8 we compare our cross sections with the low-energy
scattering data\cite{Hump,Sakitt,Kim,Kittel,Cibo,Evans}. We show the
results obtained with the full
basis of eight physical coupled states (full line),
with the isospin basis (short-dashed line) and omitting
the $\eta$ channels (long-dashed line).
The elastic cross section $K^- p \rightarrow K^- p$ is displayed in
fig. 3.
The cross section calculated with the isospin
basis is about $25 \%$ higher at low energies than the one evaluated
using the basis of physical states.
Another interesting feature is the cusp appearing around
the $K^-$ lab momentum
$p_L = 90$ MeV/c in the full basis calculation, which corresponds to
the
opening of the $\bar{K}^0 n$ channel. This cusp appears weakened and at
lower energies in the calculation with the isospin basis as a consequence
of the use of average masses for $\bar{K}$, $\pi$, $N$ and $\Sigma$.
More spectacular is the effect of omitting the $\eta \Lambda, \eta \Sigma$
channels which leads to a 60\% larger $K^- p$ elastic
cross section close to threshold and about $40 \%$ larger
around $p_L = 100$ MeV/c.

In fig. 4 we show the cross section for $K^- p \rightarrow \bar{K}^0 n$.
The results for the isospin basis  calculation and those using the
full
basis are nearly identical. Omitting the $\eta \Lambda$ and $\eta \Sigma$
channels in this case reduces the cross section in $20 \%$
around $p_L = 130$ MeV/c and above.

In fig. 5 we show the cross section for $K^- p \rightarrow \pi^0 \Lambda$.
In this case the use of the isospin basis nearly doubles the cross section
close to threshold with respect to the results with the full basis. The
effects of omitting the $\eta \Lambda, \eta \Sigma$ channels are more
moderate here and amount to an increase of about $20 \%$ in the region of
the cusp and about $10 \%$ at momenta higher than $p_L = 140$
MeV/c.

In fig. 6 we show the cross section for $K^- p \rightarrow \pi^+ \Sigma^-$.
The results using the isospin basis are about $45 \%$ larger close
to threshold
than those obtained with the full basis. The effects of omitting the $\eta$
channels are  moderate and result into an increase of the cross section of
about $12 \%$ close to threshold and a negligible change for $p_L >
100$ MeV/c.

In fig. 7 we show the cross section for $K^- p \rightarrow \pi^0 \Sigma^0$.
Although not visible in the figure, the cross section at
energies close to threshold using the isospin basis is about
$25 \%$ higher than the one obtained with the full basis. Omitting the $\eta$
channels increases the cross section in about 60\% close to threshold and in
about $30 \%$ at $p_L \simeq 100$ MeV/c.

Finally, in fig. 8 we show the results for the
$K^- p \rightarrow \pi^- \Sigma^+$ reaction. The cross sections at threshold
with the isospin and the full bases are similar, but the latter results show
a very pronounced cusp around $p_L = 90$ MeV/c corresponding to the
opening
of the $\bar{K}^0 n$ channel. This cusp is shifted to lower energies and is
less aparent in the case of the isospin basis. The omission of the $\eta$
channels has in this case a spectacular effect. The cross section is
multiplied by a factor of nearly three close to threshold when the $\eta$
channels are omitted. As a consequence the threshold ratio
$\gamma$ is very sensitive to the $\eta$ channels as is evident from
the results in Table VI. Around $p_L = 100 - 150$ MeV/c the cross
section omitting the
$\eta$ channels is about twice as large as the full calculation.

One of the novel findings of the present work is that the inclusion of the
$\eta$ channels is very important and allows one to obtain a good
reproduction of the data by means of the lowest order Lagrangian alone
using a cut off, $q_{max}$, and changing $f$ moderatly from the $f_{\pi}$
value of the meson-meson interaction.

In fig. 9, following the parallelism with the work of \cite{Kai95}, we show
the amplitudes for $K^- p \rightarrow K^- p$ and $K^- n \rightarrow K^- n$
calculated with the full basis of physical states and including the
$\eta$
channels. The results are similar to those obtained in \cite{Kai95}.

In Table VII we show the scattering lengths for $K^- p$ and $K^- n$
calculated with the three methods. We observe that  isospin breaking effects in the $K^- n$ amplitude, as well
as those omitting the $\eta$ channel, are moderate in this case. We shoud
note that this is a $T = 1$ channel where the $\Lambda (1405)$ resonance
is not present. However, the $K^- p$ amplitude, which is affected by the
presence of the resonance, shows a larger sensitivity to isospin breaking
effects and the $\eta$ channels.

The $K^- p$ scattering length is
also in good agreement with the one obtained in \cite{Kai95}.
However, the results obtained
with the isospin basis are closer to those obtained in the full basis in
our case, while in \cite{Kai95} $Re \, (a)$ is about a factor two
smaller
when average masses for $K$ and $N$ are used.

Our results for the $K^- p$ scattering length are essentially in agreement
with the most recent results from Kaonic hydrogen $X$ rays \cite{Miw97},
$(- 0.78 \pm 0.15 \pm 0.03) + i (0.49 \pm 0.25 \pm 0.12)$ fm, and in
qualitative agreement with the scattering length determined from scattering
data in \cite{Adm81}, $(- 0.67 + i \, 0.64)$ fm with 15\% estimated error.
These latter results are
obtained from the isospin scattering lengths determined in \cite{Adm81},
but as we can see from Table VII there are violations of isospin at the
level of $20 \%$ in these amplitudes.

The $K^- n$ scattering length is also in qualitative agreement with the
$T = 1$ value of \cite{Adm81}, $(0.37 + i \, 0.60)$ fm with also 15\%
estimated errors.

It is also worth calling the attention to the remarkable agreement of
our results for the real part of the scattering lengths with those
obtained in \cite{Adm81} from a combined dispersion relation and
M matrix analysis, $Re\,(a_{K^- p})=-0.98$ fm,
$Re\,(a_{K^- n})=0.54$ fm.

Next we look at the $S = 1$ sector. In fig. 10 we show the phase shifts in
the isospin channel $T = 1$. The agreement with
experiment\cite{brmart} is fair but the
phase shifts in absolute value are a little smaller than experiment. This
result is qualitatively similar to the one obtained in \cite{Kai95}, where
it was also shown that allowing for a $K^+p$ shorter range
parameter (larger cut off in our case)
the agreement with data improves.

On the other hand the scattering length in $T = 0$, which was zero at lowest
order (eq. (13)), becomes finite, although negligibly small, as a
consequence
of the coupling to other channels when the different masses are kept.
We obtain a value

\begin{equation}
a (S = 1, T = 0) = 2.4 \times 10^{- 7}\ {\rm fm}
\end{equation}

\noindent
which is compatible with present experimental data,
$0.02 \pm 0.04$ fm \cite{Do82}.
We also evaluate the scattering length for $K^+ N$ in $T = 1$, for which we
get

\begin{equation}
a (S = 1, T = 1) = - 0.26 \ {\rm fm}
\end{equation}

\noindent
which compares reasonably with the experimental number $ - 0.32 \pm 0.02$
fm \cite{Do82}. The discrepancy is similar to the one obtained for the
phase shifts in fig. 10.
For completeness we also show the phase shifts for $S = - 1, T = 2$ in fig.
11.

\section{Discussion and conclusions}

We have presented here a method of coupled-channel Lippmann Schwinger
equations which allows us to evaluate the $L = 0$ amplitudes and obtain
a good description of the $K^- p \rightarrow K^- p, \bar{K}^0 n,
\pi^0 \Lambda, \pi^0 \Sigma^0, \pi^+ \Sigma^-, \pi^- \Sigma^+$ cross
sections at low energies plus the properties of the $\Lambda (1405)$
resonance. The method uses as input only the lowest order chiral Lagrangian
which is used as a source of the potential in the LS equations, and a cut
off to regularize the loop integrals.

Using different argumentations we showed that in the loop evaluation only
the on-shell part of the potential was needed, which reduced the
coupled-channel
integral equations to algebraic equations. The approach is more
economical than the one of \cite{Kai95} in which it was inspired. Here one
obtains a good reproduction of the data without the need to use the
information from higher order Lagrangians. We should note that the
parameters of these Lagrangians do not have a fixed value. They depend upon
the energy scale chosen for the regularization \cite{Ga85}. In our language
this means that they depend upon the cut off $q_{max}$ which plays a similar
role to the energy scale in the dimensional regularization of \cite{Ga85}.
The success of our method using only the lowest order Lagrangian implies
that the chosen cut off minimizes the effect of the higher
order Lagrangians
in the L = 0 channel that we have studied. The same thing happened in the
meson-meson interaction in $L = 0$ \cite{OO97}. In \cite{Kai95} a form factor
is used with a range similar to our cut off, but a solution using only the
lowest order Lagrangians could not be found. Although a fair reproduction
of the cross sections and $\pi \Sigma$ mass distribution could be found,
the threshold parameters, particularly $\gamma$, were very poorly reproduced.
We have reconfirmed these findings omitting the $\eta \Lambda, \eta \Sigma$
channels in the coupled-channel system. However, and this is one of the main
findings of the present work, the situation is drastically changed when
these channels are included. The ratio $\gamma$ is increased by about a factor
2.2, coming in good agreement with the data, and an appreciable change in all
the channels is induced, particularly in the $K^- p \rightarrow K^- p,
\pi^0 \Sigma^0 $ reactions, and most specially in the $K^- p \rightarrow
\pi^- \Sigma^+ $ reaction whose cross section is reduced in about a factor
three at small energies.

As commented above, our fit to the data was done only for the threshold ratios
and the $\Lambda (1405)$ properties. This determined $f$ and $q_{max}$. The
value $f = 1.15 f_{\pi}$ obtained in the best fit lies between $f_{\pi}$ and
$f_K$ in the meson-meson interaction and appears as a reasonable renormalization
of $f_{\pi}$ in the $\bar{K} N$ sector. The cross sections were not
used for the fit. In spite of that, it is remarkable to see the agreement of
the results obtained with the data. The cross sections for the
$K^- p \rightarrow K^- p, \bar{K}^0 n, \pi^+ \Sigma^-, \pi^0 \Sigma^0$ are
in very good agreement with the data. Those for the
$K^- p \rightarrow \pi^- \Sigma^+$ are also compatible with the data within
errors, with small discrepancies in the deep region around $k =
90$ MeV/c largely influenced by a cusp effect in our case. Only the
$K^- p \rightarrow \pi^0 \Lambda$ cross section appears to
overestimate slightly the scarce available data.

The success of our approach in $\bar{K} N$ and coupled channels for $L = 0$
with the lowest order Lagrangian and a cut off does not mean that the
procedure can be generalized to all meson-nucleon channels. The richness of
this information most probably requires the use of higher order Chiral
Lagrangians, as it was the case in the meson-meson interaction when
including all different channels \cite{OOP97}. For the purpose of determining
these higher order terms, for a chosen scale of energies in the regularization
scheme (cut off in our method), a global fit to all meson-nucleon data would
have to be conducted in analogy to the work of \cite{OOP97}.

Meanwhile, the success of our scheme, wich is quite economical, could be
exploited to address problems related with the propagation of kaons in
matter, a topic which has aroused much interest lately
\cite{Bro93,WKW96,GEB92,CHL95}.

\acknowledgements

We would like to thank B. Krippa for discussions and some useful
checks. Comments and useful information from N. Kaiser are also acknowledged.
One of us, E. O., wishes to acknowledge the hospitality of the University
of Barcelona where part of this work was done. A. R. acknowledges finantial
help from the European network CHRX-CT 93-0323.
This work is also
partly supported by DGICYT contract numbers PB95-1249 and
PB96-0753.

\begin{table}
\centering
\caption{$C_{i j}$ coefficients of eq. (7). $C_{j i} = C_{i j}$.}
\vspace{0.5cm}

\begin{tabular}{c|cccccccc}
 & $K^- p$ & $\bar{K}^0 n$ & $\pi^0 \Lambda$ & $\pi^0 \Sigma^0$ &
$\eta \Lambda$ & $\eta \Sigma^0$ & $\pi^+ \Sigma^-$ & $\pi^- \Sigma^+$ \\
\hline
$K^- p$ & 2 & 1 & $\frac{\sqrt{3}}{2}$ & $\frac{1}{2}$ & $\frac{3}{2}$ &
$\frac{\sqrt{3}}{2}$ & 0 & 1 \\
$\bar{K}^0 n$ &  & 2 & $- \frac{\sqrt{3}}{2}$ & $\frac{1}{2}$ &
$\frac{3}{2}$ & $- \frac{\sqrt{3}}{2}$ & 1 & 0 \\
$\pi^0 \Lambda$ &  &  & 0 & 0 & 0 & 0 & 0 & 0 \\
$\pi^0 \Sigma^0$  &  &  &  &  0 & 0 & 0 & 2 & 2 \\
$ \eta \Lambda$ &  &  &  &  & 0 & 0 & 0 & 0 \\
$ \eta \Sigma^0$ &  &  &  &  &  &  0 & 0 & 0 \\
$ \pi^+ \Sigma^-$ &  &  &  &  &  &  & 2 & 0 \\
$\pi^- \Sigma^+ $  &  &  &  &  &  &  &  &  2
\end{tabular}
\end{table}

\begin{table}
\centering
\caption{$D_{i j}$ coefficients of eq. (9) for $T = 0$. $D_{j i}=D_{i
j}$.}
\vspace{0.5cm}

\begin{tabular}{c|ccc}
  & $\bar{K} N $ & $\pi \Sigma$ & $\eta \Lambda$ \\
\hline
$\bar{K} N$ & 3 &  $- \sqrt{\frac{3}{2}}$ & $\frac{3}{\sqrt{2}}$ \\
$\pi \Sigma$ &  &  4  &  0  \\
$ \eta \Lambda $ & &  &  0
\end{tabular}
\end{table}

\begin{table}
\centering
\caption{$F_{i j}$ coefficients of eq. (9) for $T = 1$. $F_{j i}=F_{i
j}$.}
\vspace{0.5cm}
\begin{tabular}{c|cccc}
  & $\bar{K} N $ & $\pi \Sigma$ & $\pi \Lambda$ & $\eta \Sigma$ \\
\hline
$\bar{K} N$ & 1 & $- 1$ & $ - \sqrt{\frac{3}{2}}$ & $ - \sqrt{\frac{3}{2}}$ \\
$\pi \Sigma $ & & 2 & 0 & 0 \\
$\pi \Lambda$ & &  &  0  &  0  \\
$\eta \Sigma$ & &  &   &  0  \\
\end{tabular}
\end{table}

\begin{table}
\centering
\caption{$L_{i j}$ coefficients for $S = - 1$. $L_{j i}=L_{i j}$.}
\vspace{0.5cm}

\begin{tabular}{c|cc}
  & $K^+ n $ &  $K^0 p $ \\
\hline
$K^+ n $ & $- 1$ & $- 1$ \\
$K^0 p $ & & $- 1$
\end{tabular}
\end{table}

\begin{table}
\centering
\caption{$\tilde{C}_{i j}$ coefficients for $K^- n$ and
related
channels in the $S = - 1, T = 1$ sector. $\tilde{C}_{j i} = \tilde{C}_{i j}$.}
\vspace{0.5cm}
\begin{tabular}{c|ccccc}
  &  $K^- n $ & $\pi^0 \Sigma^-$ & $\pi^- \Sigma^0$ & $\pi^- \Lambda$ &
$\eta \Sigma^-$ \\
\hline
$K^- n$ & 1  &  $\frac{1}{\sqrt{2}}$ & $ - \frac{1}{\sqrt{2}}$ &
$\sqrt{\frac{3}{2}}$ &  $\sqrt{\frac{3}{2}}$ \\
$\pi^0 \Sigma^-$ &  & 0 & $- 2$ & 0 & 0 \\
$\pi^- \Sigma^0$ &  &  &  0  &  0  &  0 \\
$\pi^- \Lambda$ &  &  &   &  0  &  0 \\
$\eta \Sigma^-$ &  &  &   &   &  0
\end{tabular}
\end{table}

\begin{table}
\centering
\caption{Threshold ratios}
\vspace{0.5cm}

\begin{tabular}{c|ccc}
  &  $\gamma$ & $R_c$ & $R_n$  \\
\hline
Isos. basis & 3.37 & 0.626  & 0.297  \\
Full basis & 2.33 & 0.640 & 0.217 \\
No $\eta$ & 1.05 & 0.649  & 0.164 \\
exp. \cite{No78,To71}& 2.36 $\pm$ 0.04 & 0.664 $\pm$ 0.011 & 0.189
$\pm$
0.015
\end{tabular}
\end{table}

\begin{table}
\centering
\caption{$K^- N$ scattering lengths}
\vspace{0.5cm}
\begin{tabular}{c|cc}
   &  $a_{K^- p}$[fm] & $a_{K^- n}$[fm] \\
\hline
Isos. basis & $- 0.85 + i 1.24$ & $0.54 + i 0.54$  \\
Full basis & $- 0.99 + i 0.97$  & $0.53 + i 0.61$  \\
No $\eta$ & $- 0.64 + i 1.66$   & $0.47 + i 0.53$ \\
\cite{Kai95} & $- 0.97 + i 1.1$ & \\
exp. \cite{Miw97} & $(-0.78\pm 0.18) + i(0.49\pm 0.37)$ & \\
exp. \cite{Adm81} & $- 0.67 + i 0.64$ & $0.37 + i 0.60$ \\
exp. $Re\,(a)$\cite{Adm81} & $- 0.98 $ & $0.54$
\end{tabular}
\end{table}

\begin{figure}
       \setlength{\unitlength}{1mm}
       \begin{picture}(100,180)
       \put(25,0){\epsfxsize=12cm \epsfbox{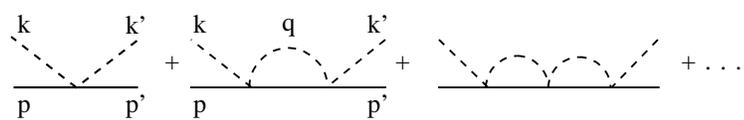}}
       \end{picture}
\caption{
Diagrammatic representation of the Lippmann Schwinger equations,
eq. (15), in $\bar{K} N$ scattering.
}
\end{figure}

\begin{figure}
       \setlength{\unitlength}{1mm}
       \begin{picture}(100,180)
       \put(25,0){\epsfxsize=12cm \epsfbox{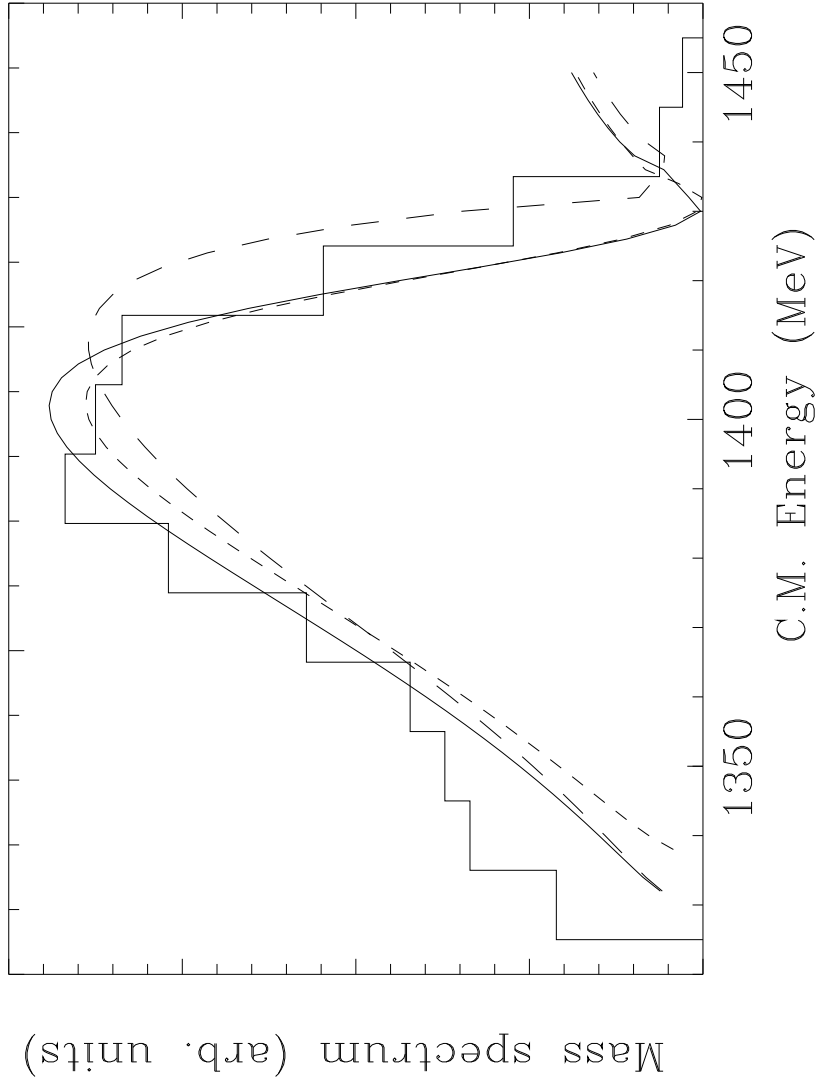}}
       \end{picture}
\caption{
The $\pi \Sigma$ mass distribution around the $\Lambda (1405)$ resonance
from eq. (27).
Short-dashed line: results in isospin
basis. Long-dashed line: results omitting the $\eta \Sigma^0, \eta
\Lambda$ channels. Full line:
results with the full basis of physical states.
Experimental data from \protect\cite{Th73}.
}
\end{figure}

\begin{figure}
       \setlength{\unitlength}{1mm}
       \begin{picture}(100,180)
       \put(25,0){\epsfxsize=12cm \epsfbox{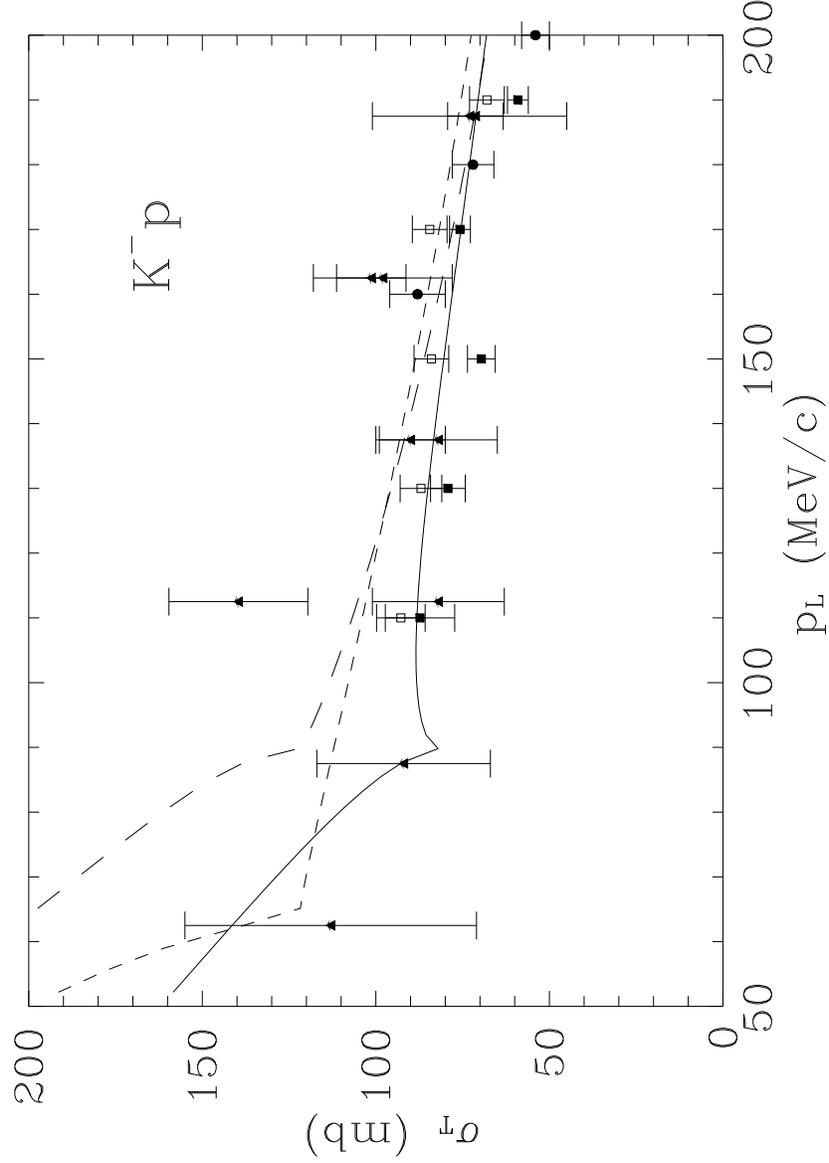}}
       \end{picture}
\caption{
$K^- p \rightarrow K^- p$ cross section as a function of the $K^-$
momentum in the lab frame.
Short-dashed line: results in isospin
basis. Long-dashed line: results omitting the $\eta \Sigma^0, \eta
\Lambda$ channels. Full line:
results with the full basis of physical states.
For Figs.
3--8, the experimental data are from:
\protect\cite{Hump} (black triangles), \protect\cite{Sakitt} (black
squares),
\protect\cite{Kim} (open squares), \protect\cite{Kittel} (open
triangles), \protect\cite{Cibo}
(black circles) and \protect\cite{Evans} (open circles).
}
\end{figure}

\begin{figure}
       \setlength{\unitlength}{1mm}
       \begin{picture}(100,180)
       \put(25,0){\epsfxsize=12cm \epsfbox{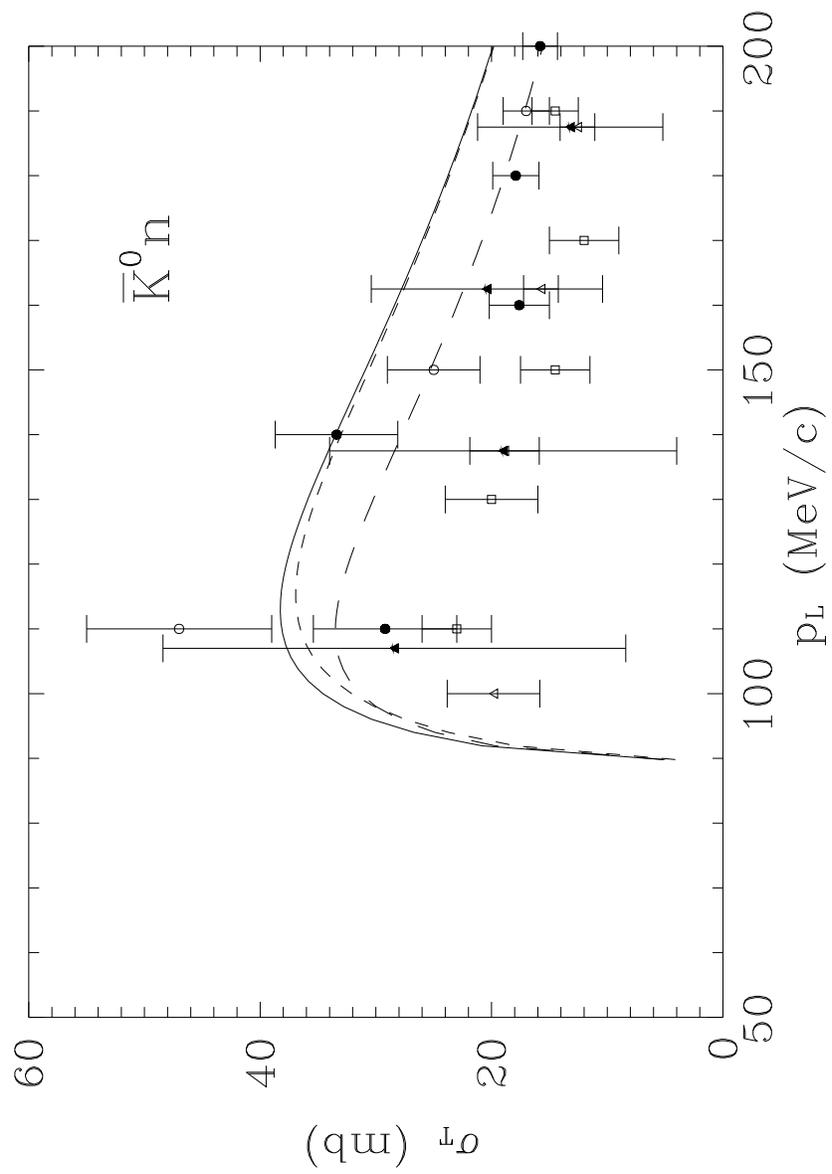}}
       \end{picture}
\caption{
Same as fig. 3 for $K^- p \rightarrow \bar{K}^0 n$
}
\end{figure}

\begin{figure}
       \setlength{\unitlength}{1mm}
       \begin{picture}(100,180)
       \put(25,0){\epsfxsize=12cm \epsfbox{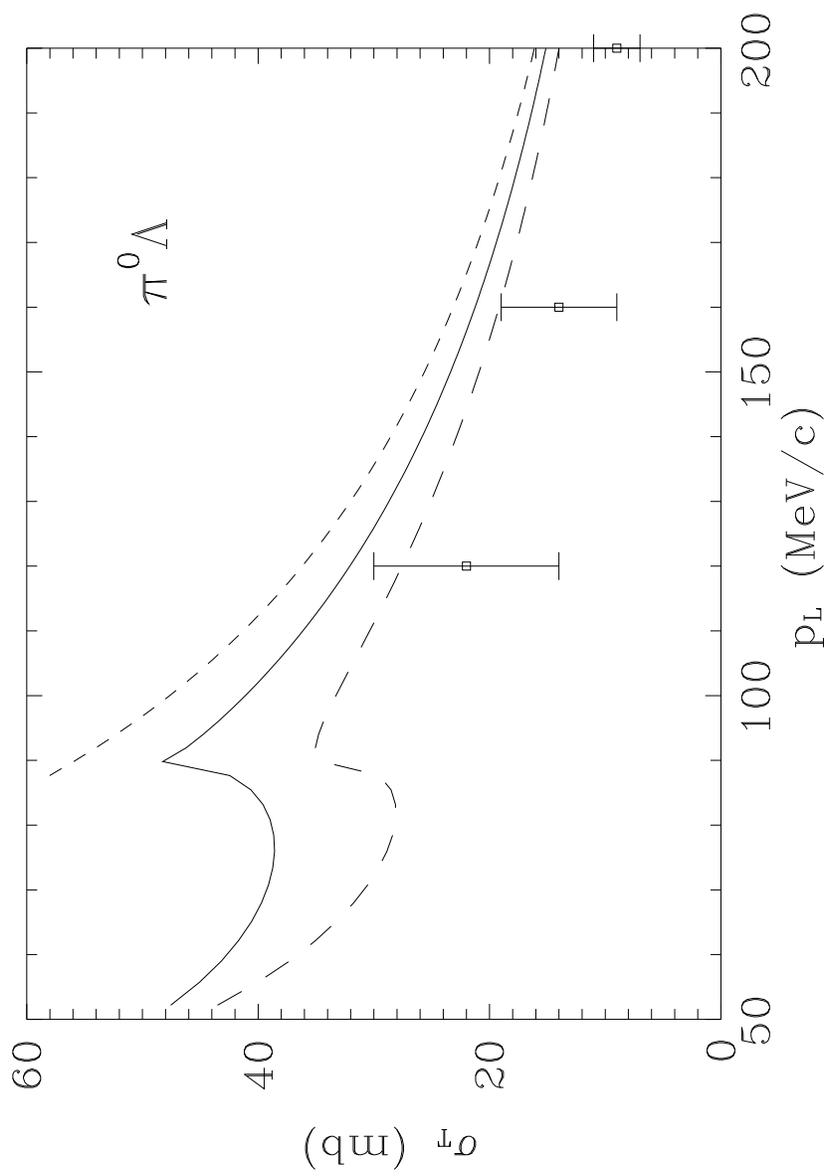}}
       \end{picture}
\caption{
Same as fig. 3 for $K^- p \rightarrow \pi^0 \Lambda$
}
\end{figure}

\begin{figure}
       \setlength{\unitlength}{1mm}
       \begin{picture}(100,180)
       \put(25,0){\epsfxsize=12cm \epsfbox{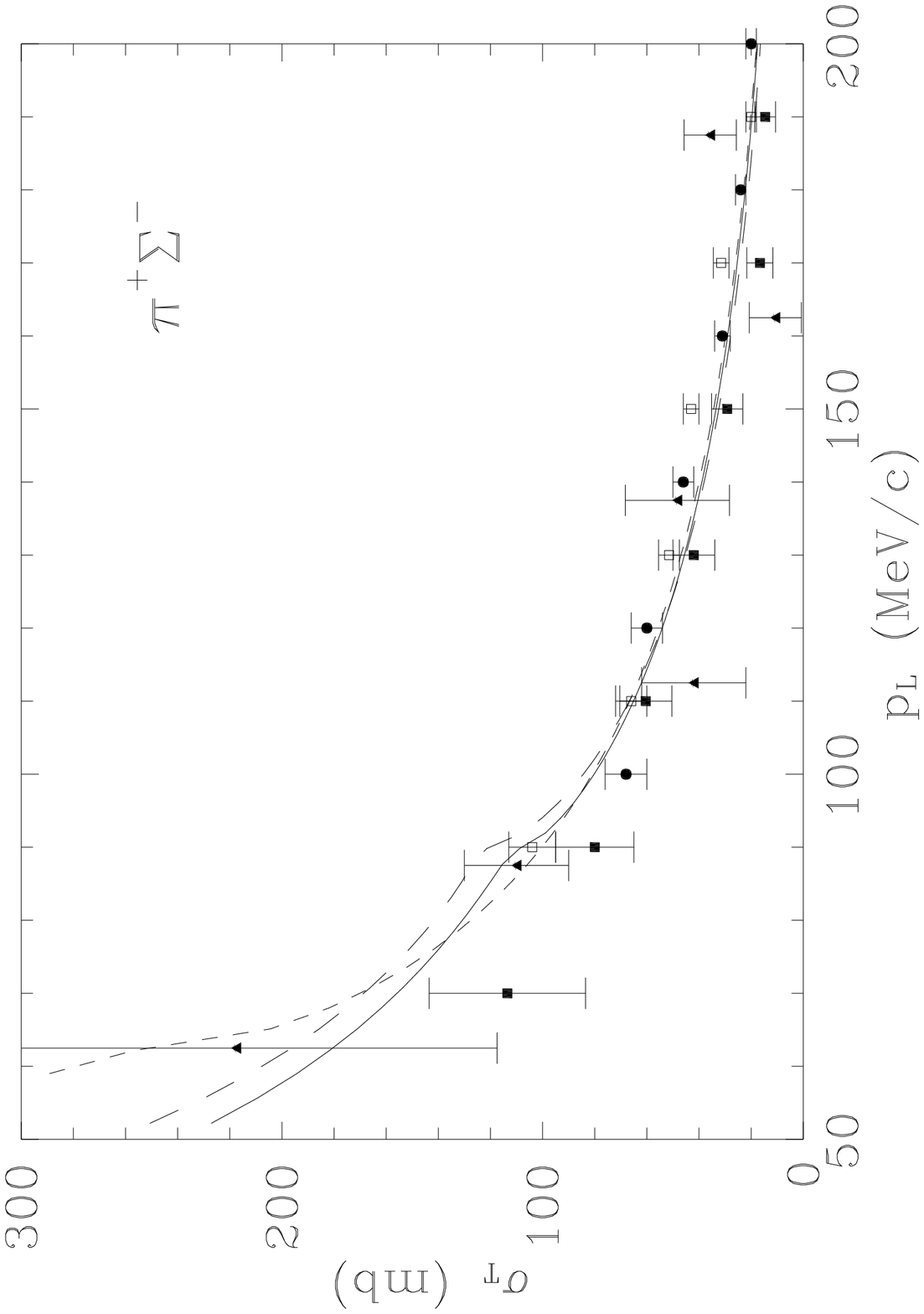}}
       \end{picture}
\caption{
Same as fig. 3 for $K^- p \rightarrow \pi^+ \Sigma^-$
}
\end{figure}

\begin{figure}
       \setlength{\unitlength}{1mm}
       \begin{picture}(100,180)
       \put(25,0){\epsfxsize=12cm \epsfbox{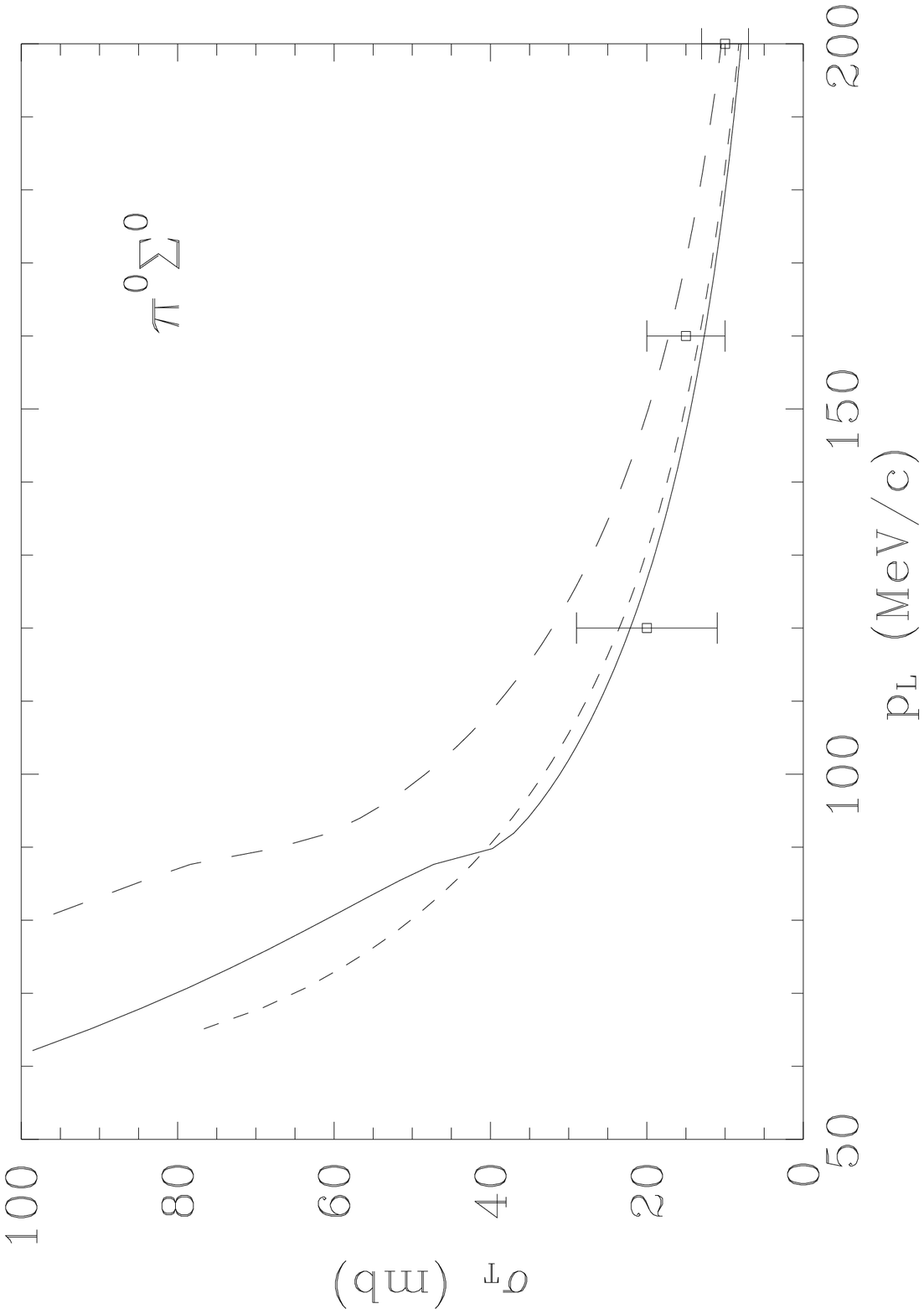}}
       \end{picture}
\caption{
Same as fig. 3 for $K^- p \rightarrow \pi^0 \Sigma^0$
}
\end{figure}

\begin{figure}
       \setlength{\unitlength}{1mm}
       \begin{picture}(100,180)
       \put(25,0){\epsfxsize=12cm \epsfbox{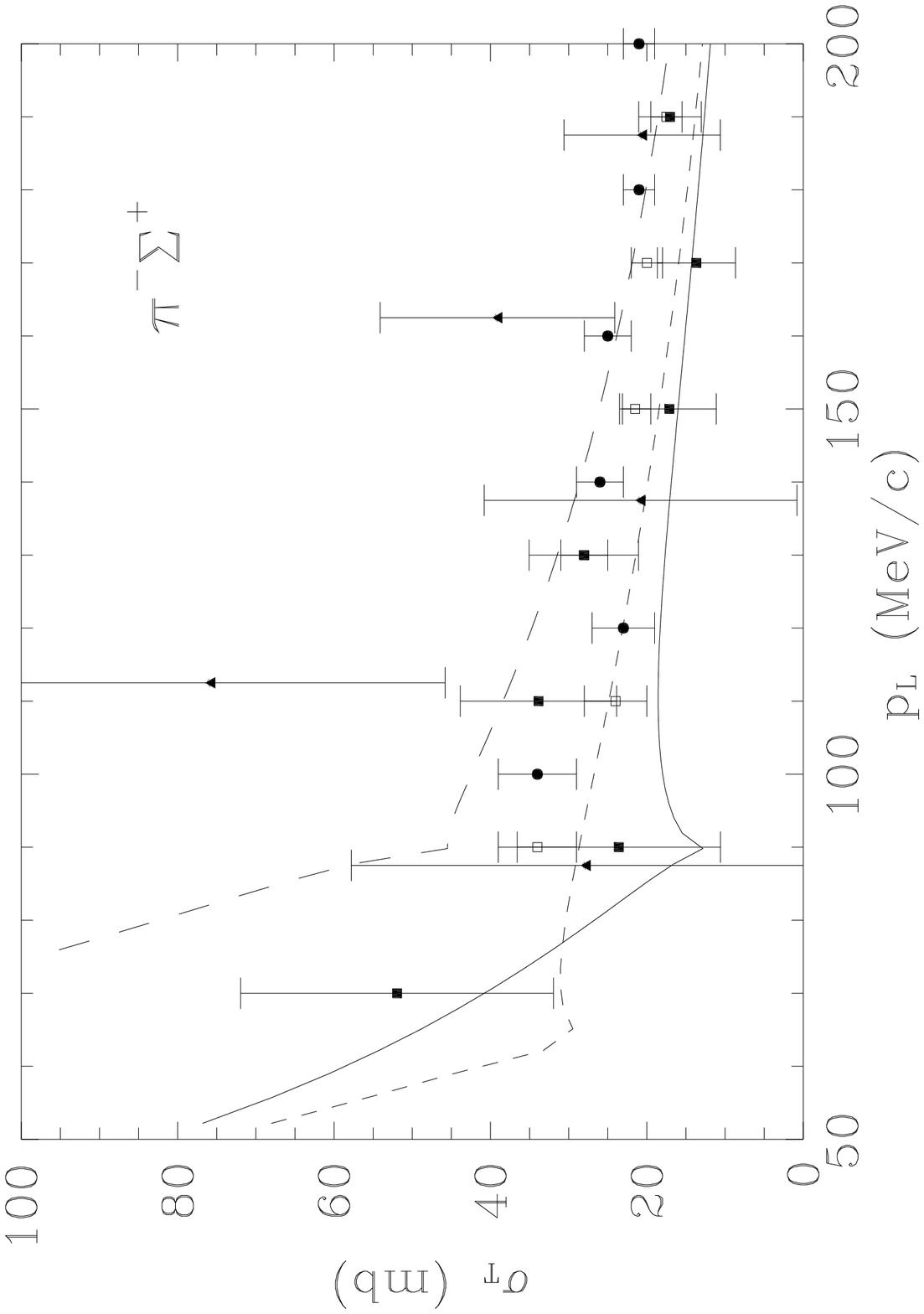}}
       \end{picture}
\caption{
Same as fig. 3 for $K^- p \rightarrow \pi^- \Sigma^+ $
}
\end{figure}

\begin{figure}
       \setlength{\unitlength}{1mm}
       \begin{picture}(100,180)
       \put(25,0){\epsfxsize=12cm \epsfbox{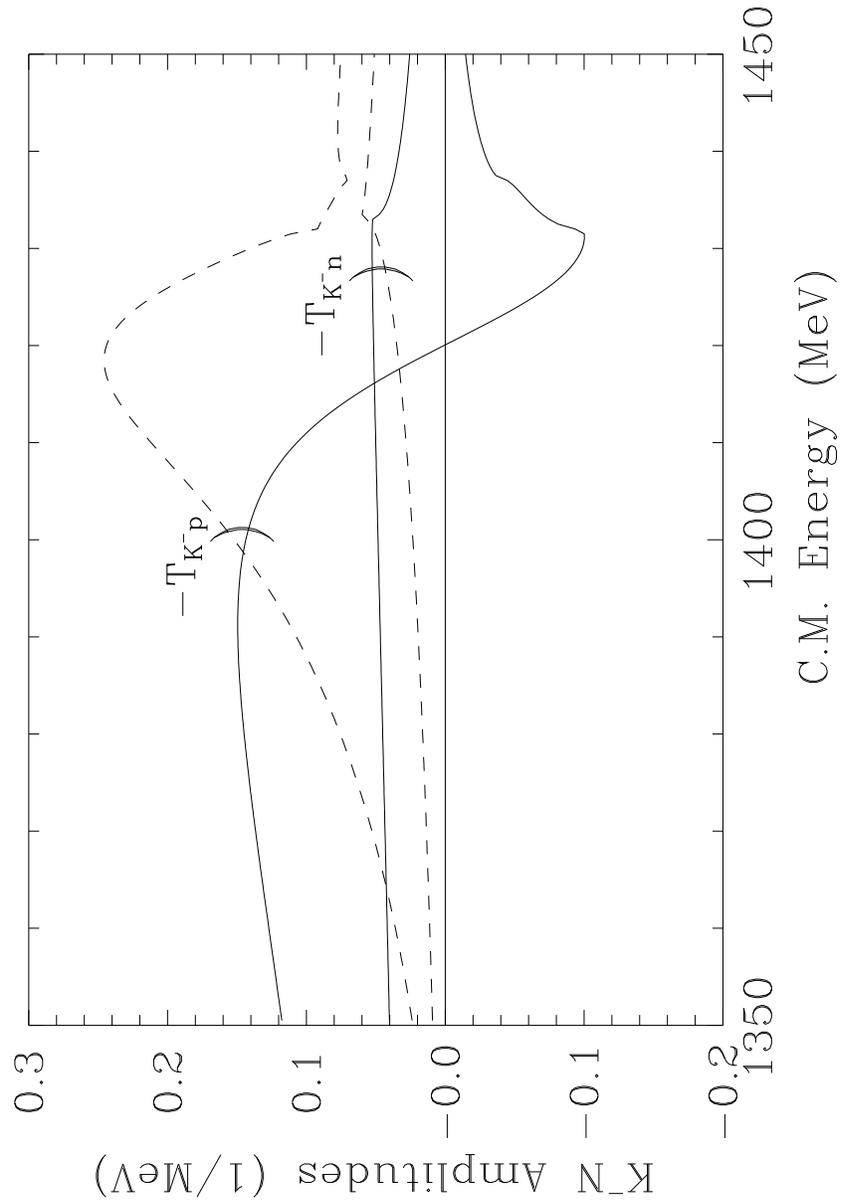}}
       \end{picture}
\caption{
Scattering amplitudes for $K^- p \rightarrow K^- p$ and
$K^- n \rightarrow K^- n$ around and below the $K^- N$ threshold.
Solid lines: real part. Dashed lines: imaginary part.
}
\end{figure}

\begin{figure}
       \setlength{\unitlength}{1mm}
       \begin{picture}(100,180)
       \put(25,0){\epsfxsize=12cm \epsfbox{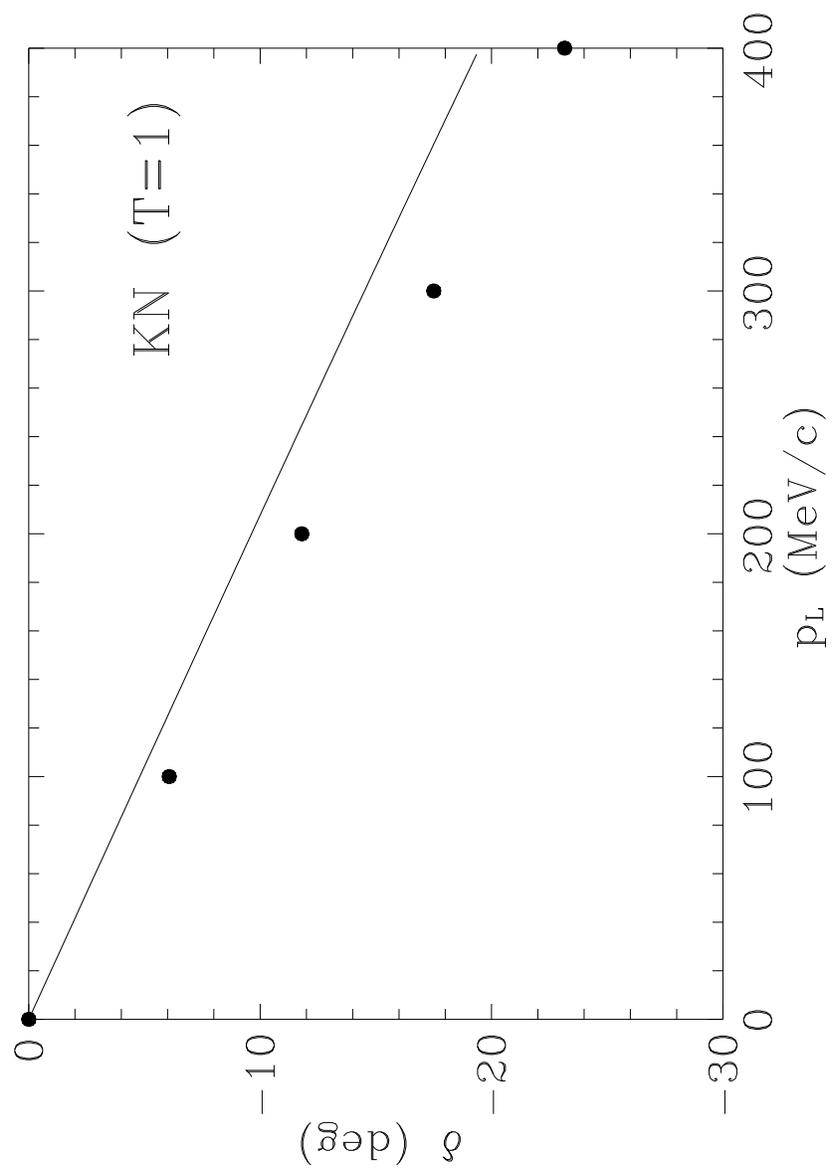}}
       \end{picture}
\caption{
S-wave phase shifts for $K N$ in $T = 1$ as a function of the kaon
lab momentum.
}
\end{figure}

\begin{figure}
       \setlength{\unitlength}{1mm}
       \begin{picture}(100,180)
       \put(25,0){\epsfxsize=12cm \epsfbox{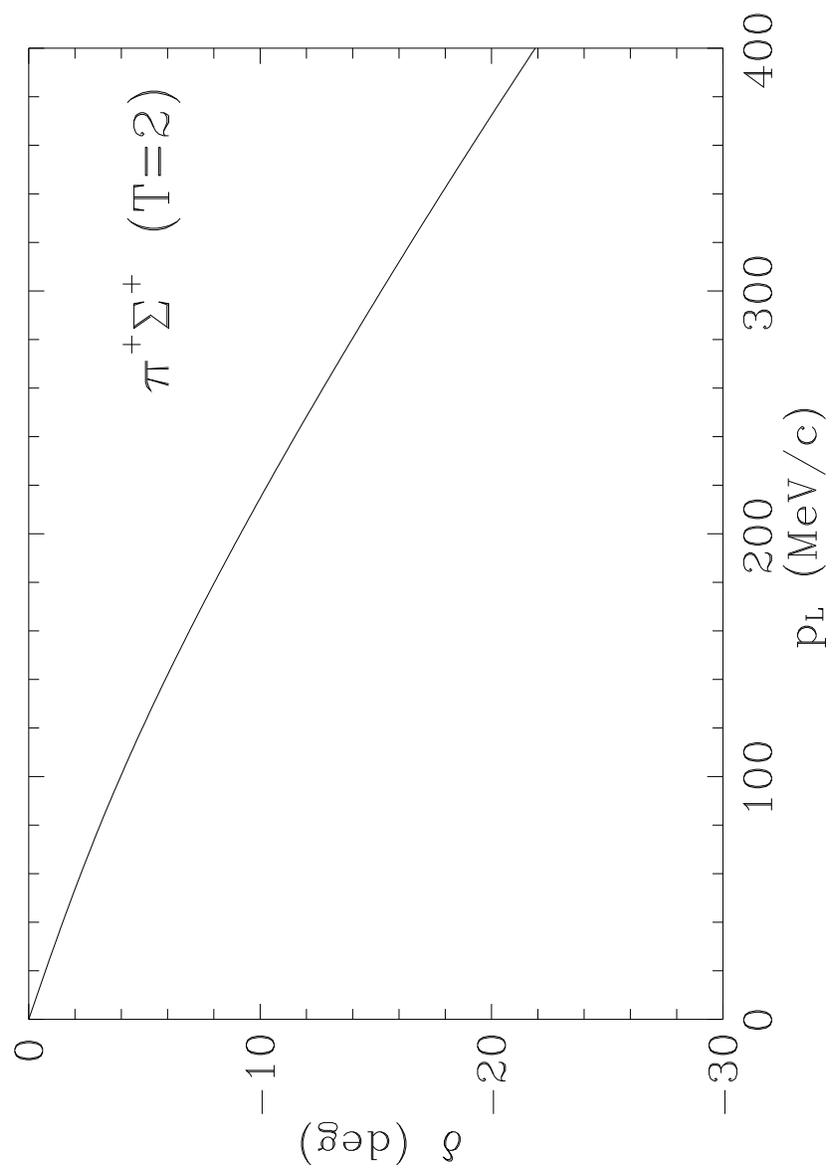}}
       \end{picture}
\caption{
S-wave phase shifts for $\pi^+ \Sigma^+$ as a function of the pion
lab momentum.
}
\end{figure}

\end{document}